\documentclass[%
 preprint,
superscriptaddress,
 amsmath,amssymb,
 aps,
 prl,
]{revtex4-2}

\usepackage{graphicx}
\usepackage{dcolumn}
\usepackage{bm}
\usepackage[mathlines]{lineno}
\usepackage{braket}
\usepackage[dvipsnames]{xcolor}
\usepackage{amsmath,amssymb,amsfonts} 
\usepackage{float}
\usepackage{bbm}
\usepackage{nicefrac}
\usepackage{wasysym}
\usepackage[normalem]{ulem}

\usepackage{hyperref}
\hypersetup{
	colorlinks=true,
	linkcolor=blue,
	filecolor=blue,
	urlcolor=blue,
	citecolor=blue,
}

\def\beq{\begin{equation}}
\def\eeq{\end{equation}}
\def\bea{\begin{eqnarray}}
\def\eea{\end{eqnarray}}

\def\romanI{\textnormal{\MakeUppercase{\romannumeral 1}}}
\def\romanII{\textnormal{\MakeUppercase{\romannumeral 2}}}
\def\kk{{\bm k}}
\def\rmO{{\rm O}}
\def\rmS{{\rm S}}

\begin{document}

\title{Intrinsic Orbital Origin for the Chirality-Dependent Nonlinear Planar Hall Effect of Topological Nodal Fermions in Chiral Crystals}

\author{Mingxiang Pan}
\affiliation{School of Physics, Peking University, Beijing 100871, China}

\author{Hui Zeng}
\affiliation{State Key Laboratory of Low-Dimensional Quantum Physics, Department of Physics, Tsinghua University, Beijing 100084, China}

\author{Erqing Wang}
\affiliation{School of Physics, Peking University, Beijing 100871, China}

\author{Huaqing Huang}
\email[Corresponding author: ]{huaqing.huang@pku.edu.cn}
\affiliation{School of Physics, Peking University, Beijing 100871, China}
\affiliation{Collaborative Innovation Center of Quantum Matter, Beijing 100871, China}
\affiliation{Center for High Energy Physics, Peking University, Beijing 100871, China}

\date{\today}

\begin{abstract}
Topological semimetals in chiral crystals, which possess both structural handedness and band crossings (or nodes) with topological chiral charge, exhibit many exotic physical properties. Here we demonstrate that the structural and electronic chirality of these systems can endow them with another fascinating phenomenon---the intrinsic nonlinear planar Hall effect (INPHE), which is prominent around the nodes and reverses sign upon chirality reversal in opposite enantiomers. Taking chiral tellurium as an example, we reveal an intrinsic orbital mechanism, which manifests diverging orbital magnetic moments with hedgehog-like textures around nodes and, therefore, generates a dominant contribution to the INPHE that is proportional to the topological charge. Furthermore, we show that multifold fermions in topological chiral semimetals with B20 structures (e.g., CoSi and PtAl) induce a giant INPHE conductivity reaching the order of $1\sim 10\; \mathrm{A}\cdot\mathrm{V}^{-2}\cdot\mathrm{T}^{-1}$, which is detectable in experiments. Our study not only relates nonlinear transport to band topology and enantiomer recognition but also offers a new way to explore the exotic physical properties associated with unconventional chiral fermions.
\end{abstract}

\maketitle

Chiral crystals which possess a well-defined handedness due to the lack of inversion and mirror symmetries, exhibit numerous fascinating non-reciprocal phenomena \cite{Nagaosa2018Nonreciprocal, annurev-conmatphys-060220-100347, annurev-conmatphys-032822-033734, yan2023structural}, such as electric magnetochiral anisotropy \cite{Nat2022_EMCA_CsV3Sv5, PhysRevLett.87.236602, PhysRevB.99.245153, NatCom_EMCA_2014} and chirality-induced spin selectivity \cite{CISS_jz300793y, CISS_NRC2019, PhysRevLett.124.166602, PhysRevLett.127.126602}.
In topological chiral semimetals, such as tellurium (Te) \cite{PhysRevLett.114.206401, PhysRevLett.125.216402, PhysRevB.95.125204, PhysRevLett.124.136402}
and B20 compounds (e.g., CoSi and PtGa) \cite{Change2018TQPchiral,nc2019_chiral_CoSi_RhSi_PtAl}, in addition to structural chirality, a definite electronic chirality can be assigned to electron wavefunctions at point-like two- or multi-band crossings (nodes) of the quasiparticle dispersions, which are known as Weyl or multifold fermions. These nodal fermions are characterized by an integer topological chiral charge $\mathcal{C}$, which is the Chern number for any surface enclosing the nodal point in momentum space \cite{science.aaz3480,nc2020_maxChern_PtGa}. The electronic chirality, which reverses signs in chiral variants with opposite handednesses (enantiomers), also results in a plethora of novel phenomena, such as long Fermi-arc surface states \cite{Nat2019.chiral_FermiArc_RhSi, Nat2019.HongDing_FermiArc_CoSi, Natphy2019.YulinChen_FermiArc_AlPt}, unconventional radial spin textures \cite{PhysRevLett.114.206401, PhysRevB.95.125204,PhysRevLett.125.216402, PhysRevLett.124.136402, PhysRevMaterials.7.014204, adfm.202208023},
and most notably the quantized circular photogalvanic effect whose magnitude associated with optical transitions near a nodal point is proportional to its topological charge \cite{quantized_cpge2017moore, sciadv.aba0509, nc2021_cpge_CoSi, nc2022_cpge_Te, acs.nanolett.3c00780}.
In this Letter, we demonstrate another novel chirality-dependent phenomenon---the nonlinear planar Hall effect (NPHE), which strongly enhances around nodal points and reverses sign in opposite-handed enantiomers. We further reveal a previously overlooked intrinsic orbital origin of this effect, which is endowed by the electronic chirality.

The NPHE generates a nonlinear Hall response current ($j_{\rm{H}}\propto E^2B$) by a driving $E$ field under an in-plane magnetic $B$ field, and therefore, can be regarded as the magnetic perturbated response of the recently-discovered nonlinear anomalous Hall effect \cite{PhysRevLett.115.216806, NAHE_nature.565.337, NAHE_in_WTe2, gao_field_2014, PhysRevLett.127.277201, PhysRevLett.127.277202, PhysRevLett.131.056401, science.adf1506, Nature2023_QM_NAHE}.
The NPHE was experimentally observed at the two-dimensional (2D) surface of topological insulators \cite{PhysRevLett.123.016801} and recently were reported in chiral semimetals Te and CoSi with a clear dependence on structural chirality \cite{acs.nanolett.3c01797, NPHE_CoSi}. Previous theoretical studies on this effect mostly focus on the extrinsic mechanisms that depend on scattering and scale as $\tau^2$ in electron relaxation time $\tau$ \cite{PhysRevB.103.155415, PhysRevB.108.L241104, PhysRevResearch.3.L012006, PhysRevB.108.075155}. More recently, a $\tau$-independent intrinsic contribution was proposed by considering the Zeeman-type coupling between the in-plane $B$ field and electron spin in 2D noncentrosymmetric materials \cite{PhysRevLett.130.126303}.
However, the orbital degree of freedom which also couples to the $B$ field and may result in NPHE, has been relatively unexplored.
Amongst, the orbital effect in chiral crystals that exhibit unique orbital textures has remained elusive \cite{pnas.2305541120}. Moreover, despite the success of the experiments, the underlying microscopic mechanism of the chirality-dependence of NPHE in chiral semimetals has not been fully recognized.

In this Letter, we demonstrate a direct connection between the NPHE and nodal fermions in chiral crystals and reveal an intrinsic orbital mechanism that dominates the effect. Taking Te as a prototypical example of topological chiral semimetals, we show that the intrinsic NPHE (INPHE) is strongly enhanced around the Weyl point (WP) and reverses sign in opposite-handed enantiomers. Based on a generic $k\cdot p$ model, we analytically demonstrate that the orbital magnetic moment around WPs exhibits a hedgehog-like texture that can be characterized by electronic chirality. Consequently, the pronounced INPHE is mainly attributed to the enhanced orbital magnetic moment surrounding WPs and proportional to the topological chiral charge. In addition, we show that multifold fermions in topological chiral semimetals, including CoSi, RhSi, RhSn, PtGa, and PtAl, can also result in chirality-dependent INPHE with even larger response amplitudes associated with their high Chern numbers.

\begin{figure}
    \hypertarget{fig1}{}
    \centering
    \includegraphics[width=1\columnwidth]{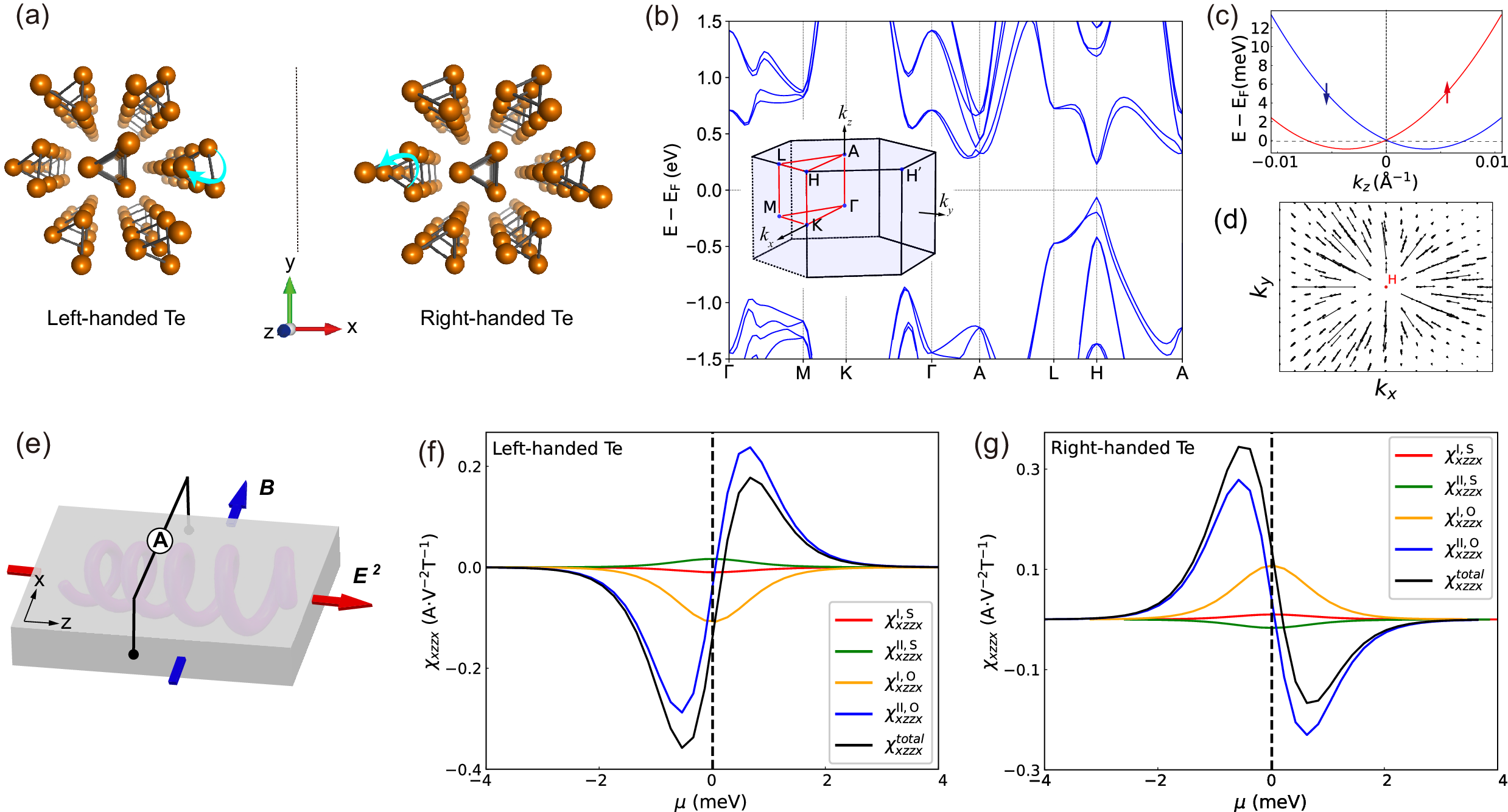}
    \caption{(a) Crystal structure of left-handed and right-handed Te from the top view. (b) The band structure of right-handed Te. The insert shows the Brillouin zone. (c) The enlarged conduction bands near the WP along $k_z$. Red and blue represent positive and negative z-components of spin, respectively. (d), Berry curvature vector field of right-handed Te around H in $k_x$-$k_y$ plane. (e) A sketch of the experimental setup to measure $\chi_{xzzx}$. (f,g) $\chi_{xzzx}$ versus the chemical potential $\mu$ for left- and right-handed Te.}
    \label{fig:band}
\end{figure}

\textit{Weyl points in chiral Te.}---The chiral crystal structure of Te consists of weakly interacting atomic spiral chains, as shown in Fig.~\ref{fig:band}(a). The left- and right-handed Te belongs to space groups $P3_221$ (No. 154, $D_3^6$) and $P3_121$ (No. 152, $D_3^4$) depending on the handedness of the screw axis along the z-axis. Our first-principles calculations \footnote{\label{fn}See Supplemental Material at http://link.aps.org/supplemental/xxx, for more details about the analytical derivation of the INPHE based on generic $k\cdot p$ models, details of first-principles calculation methods, and numerical results of INPHE for chiral Te and B20 compounds Cosi, RhSi, RhSn, PtGa, and PtAl, which include Refs.~\cite{ JPSJ.28.36, bir1974symmetry, PhysRevLett.116.247201, PhysRevB.108.L201114, PhysRevLett.129.186801, KRESSE199615, PhysRevLett.77.3865, Heyd2003HybridFB, PhysRevLett.102.073005, doi:10.1021/acs.nanolett.7b01029, RevModPhys.84.1419}.}
\cite{ JPSJ.28.36, bir1974symmetry, PhysRevLett.116.247201, PhysRevB.108.L201114, PhysRevLett.129.186801, KRESSE199615, PhysRevLett.77.3865, Heyd2003HybridFB, PhysRevLett.102.073005, doi:10.1021/acs.nanolett.7b01029, RevModPhys.84.1419}
show that both the conduction band minimum (CBM) and valence band maximum (VBM) are close to but slightly off the BZ corners (H and H') point with a band gap of about 0.3 eV (see Fig.~\ref{fig:band}(b). As shown in Fig.~\ref{fig:band}(d), spin-orbit coupling lifts the spin degeneracy at the bottom of the conduction band and forms a WP at H and H' due to the lack of inversion and mirror symmetry \cite{PhysRevLett.114.206401, PhysRevB.95.125204,PhysRevLett.125.216402, PhysRevLett.124.136402, PhysRevMaterials.7.014204, PhysRevResearch.3.023111, PhysRevB.97.035158}.
More importantly, the WPs in the right-handed Te behave as sources of the Berry curvature field in $\kk$-space and the radial spin textures are outward [see Figs.~\ref{fig:band}(c,d)], indicating their electronic chirality of $\mathcal{C}=+1$. In contrast, the WPs in the left-handed Te have the opposite chirality $\mathcal{C}=-1$ and inward radial spin texture. In both crystals, the chiralities are the same between WPs at H and H', which are linked by the time-reversal symmetry. However, it does not violate the Nielsen–Ninomiya no-go theorem~\cite{NIELSEN1981219} because their oppositely chiral charged partners exist above in energy. This leaves the WPs near CBM an ideal playground to study the chirality-dependent INPHE. Given that the WPs are about 2.5 meV above the CBM, hereafter we set the Fermi energy $E_F=0$ at the WP for \textit{n}-doped Te, which has been experimentally realized by an external gating of exfoliated flakes or the hydrothermal growth with dielectric doping technique \cite{PhysRevLett.125.216402,qiu2020QHE_n-type_Te, Wang2018n-type,acs.nanolett.1c01705}.

\textit{Intrinsic nonlinear planar Hall effect}.---The INPHE conductivity $\chi^{\rm int}$ defined by $j_a^{\rm int} = \chi_{abcd}^{\rm int}E_bE_cB_d$ ($a,b,c,d$ are Cartesian indices) can be expressed in terms of band quantities as (see Supplemental Material (SM) for more details \footnotemark[\value{footnote}]), \begin{eqnarray}
        \chi_{abcd}^{\rm int} &=& \sum_n\int_{\rm BZ} \frac{d^3\boldsymbol{\mathit{k}}}{(2\pi)^3}f^{\prime}(\epsilon_n) [\alpha_{abcd}^{n,\romanI}(\boldsymbol{\mathit{k}})+\alpha_{abcd}^{n,\romanII}(\boldsymbol{\mathit{k}})], \label{chi_int}\\
        \alpha_{abcd}^{n,\romanI} &=& v_a^n\Lambda_{bcd}^n-v_b^n\Lambda_{acd}^n,\label{IO}\\
        \alpha_{abcd}^{n,\romanII} &=& (\partial_aG_{bc}^n-\partial_bG_{ac}^n)\mathcal{M}_d^n\label{IIO},
\end{eqnarray}
where $f^{\prime}$ is the derivative of the equilibrium Fermi distribution function $f(\epsilon; T,\mu)$ [$T$ and $\mu$ are omitted in Eq.~\eqref{chi_int}]. The superscript I and II represent the contributions from the magnetic susceptibility $\Lambda^n_{abc}$ of Berry connection polarizability (BCP) $G^n_{ab}$ \cite{PhysRevLett.112.166601, PhysRevB.105.045118} and from the combination of the BCP dipole and magnetic moment $\mathcal{M}_d^n$. Explicitly,
\begin{equation}
    G_{ab}^n(\boldsymbol{\mathit{k}})=2\textnormal{Re}\sum_{m\neq n}\frac{v_a^{nm}v_b^{mn}}{(\epsilon_n-\epsilon_m)^3}\label{BCP},
\end{equation}
and $\Lambda_{abc}^{n,\textnormal{S(O)}}=\partial G_{ab}^n(\boldsymbol{\mathit{k}})/\partial B_c$
depend on the interband spin and orbital magnetic moments $\bm{\mathcal{M}}^{mn,\textnormal{S}}=-g\mu_B\boldsymbol{\mathit{s}}^{mn}$ and $\bm{\mathcal{M}}^{mn,\textnormal{O}}=\sum_{l\neq n}(\boldsymbol{\mathit{v}}^{ml}+\delta^{lm}\boldsymbol{\mathit{v}}^n)\times \mathcal{A}^{ln}/2$, with $g$ the $g$ factor and $\mu_B$ the Bohr magneton. Here $\mathcal{A}^{ln}$ is the unperturbed interband Berry connection and $\boldsymbol{\mathit{s}}^{mn}(\boldsymbol{\mathit{v}}^{ml})$ is the spin (velocity) matrix element \cite{PhysRevLett.130.126303, PhysRevB.108.075155}.
Therefore, both parts I and II of $\chi^{\rm int}_{abcd}$ can be further divided into spin and orbital contributions.

For usual planar Hall measurements in the \textit{x}-\textit{z} plane of chiral Te \cite{acs.nanolett.3c01797} [see Fig.~\ref{fig:band}(e)], only two components $\chi_{xzzx}$ and $\chi_{zxxz}$ are allowed by the $C_{2x}$ symmetry. For general cases where the applied electric and magnetic fields are not aligned with crystal axes but make polar angles $\theta$ and $\phi$ from the \textit{z}-axis, i.e., $\bm{E}=E(\sin\theta,0,\cos\theta)$ and $\bm{B}=B(\sin\phi,0,\cos\phi)$, the angle-dependent INPHE conductivity is
\begin{equation}
    \chi_{\rm H}^{\rm int}(\theta,\phi)=\chi_{xzzx}\sin\phi\cos\theta-\chi_{zxxz}\cos\phi\sin\theta.\label{angle}
\end{equation}
Here we focus on $\chi_{xzzx}$ to describe the INPHE in chiral Te and defer
$\chi_{zxxz}$ which exhibits similar behavior in Fig. S2 of SM \footnotemark[\value{footnote}].
For \textit{n}-type Te with opposite handednesses, $\chi_{xzzx}^{\rm int}$ rapidly develops peaks on the order of $10^{-1}\mathrm{A}\cdot\mathrm{V}^{-2}\cdot\mathrm{T}^{-1}$ as $\mu$ approaches CBM, and undergoes a sign change when $\mu$ crosses the WPs, as shown in Fig.~\ref{fig:band}(f,g). Remarkably, the left- and right-handed Te exhibit opposite $\chi_{xzzx}^{\rm int}$, implying the chirality-dependence of INPHE in chiral Te. Remarkably, our results are consistent with the recent experiment on chiral Te which reported the gate-tunable and chirality-dependent NPHE with cosine angular dependences on the $B$ field \cite{acs.nanolett.3c01797}. In addition, we estimate extrinsic contributions that depend on $\tau^2$ and found that their effects are negligible, indicating that INPHE is the primary mechanism (see Sec. IV in SM~\footnotemark[\value{footnote}]). It is also worth noting that $\chi^{\rm int}$ mainly comes from part II of the orbital contribution $\chi_{xzzx}^{\romanII,\rmO}$ which originates from BCP dipole and orbital magnetic moment [see Eq.~\eqref{IIO}]. This is different from the conventional spin-dominated case \cite{PhysRevLett.130.126303}, indicating a distinct physical mechanism, as we will discuss later.

\begin{figure}
    \hypertarget{dG}{}
    \centering
    \includegraphics[width=1\columnwidth]{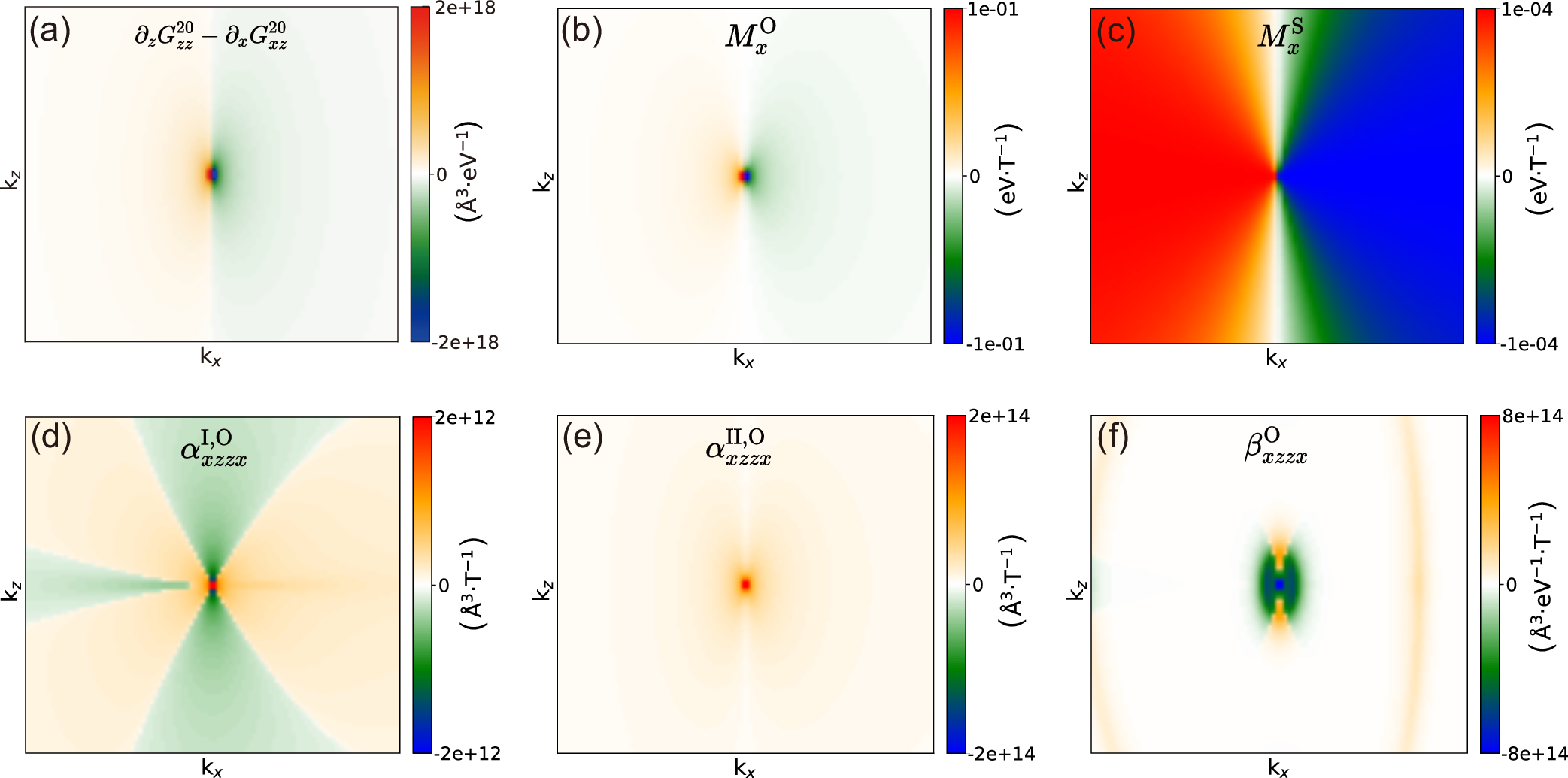}
    \caption{\label{figBCP}(a) The BCP dipole $\lambda_{xzz}^n(\kk)=\partial_xG^n_{zz}-\partial_zG^n_{xz}$, (b,c) orbital and spin magnetic moment $\mathcal{M}_x^\rmO(\kk),\mathcal{M}_x^\rmS(\kk)$, (d,e) $\alpha_{xzzx}^{n,\romanI,\rmO}(\kk)$ and $\alpha_{xzzx}^{n,\romanI,\rmO}(\kk)$ for the upper conduction band of right-handed Te. (f) The orbital part of the integrand $\beta_{xzzx}^\rmO(\kk)=\sum_n f^\prime(\epsilon_n)(\alpha^{n,\romanI,\rmO}+\alpha^{n,\romanII,\rmO})$ of Eq.~\eqref{chi_int} at 4 meV above the WP for the right-handed Te. These quantities are shown in a patch ($[-2.2,2.2]\times[-0.8,0.8]$ in units of $10^{-3}$\AA$^{-1}$) of the $k_x$-$k_z$ plane around H.
    }
\end{figure}

To gain further insight into the dominant component $\chi_{xzzx}^{\romanII,\rmO}$,
we analyze the $\kk$-space distribution of relevant band geometric quantities, including the BCP dipole $\lambda^n_{xzz}=(\partial_x G_{zz}^n-\partial_z G_{xz}^n)$, magnetic moment $\mathcal{M}_x^{n}$, and $\alpha_{xzzx}^{\romanI(\romanII),\rmO}$ for the upper conduction bands. As shown in Fig.~\ref{figBCP}, these quantities are pronounced around H where the energy difference between the two bands ($\Delta\epsilon$) is small. This is because similar to Berry curvature [Fig.~\ref{fig:band}(c)], these quantities are endowed by the interband coherence and scale as $1/(\Delta \epsilon)^\eta$ with $\eta$ being some integers (see SM \footnotemark[\value{footnote}]). Since both $\lambda^n_{xzz}$ and $\mathcal{M}_x^{n,\rmO}$ are odd functions of $k_x$ [Fig.~\ref{figBCP}(a,b)], the resultant $\alpha_{xzzx}^{\romanII,\rmO}$ exhibit a single negative peak for the upper band [Fig.~\ref{figBCP}(e)]. Moreover, we find that the lower conduction band has an opposite peak of $\alpha_{xzzx}^{\romanII,\rmO}$ (see Fig. S5 in SM \footnotemark[\value{footnote}]), indicating a sign reversal of $\chi_{xzzx}^{\romanII,\rmO}$ when $\mu$ passes the WP. In contrast,
$\alpha_{xzzx}^{\romanI,\rmO}$ display similar quadrupole distributions for both bands, resulting in small uncompensated contributions to the conductivity and without $\mu$-dependent sign reversal. In addition, one observes that the spin magnetic moment $\mathcal{M}_x^{n,\rmS}$ is orders of magnitude smaller than $\mathcal{M}_x^{n,\rmO}$ [see Fig.~\ref{figBCP}(b,c)], hence the spin contributions $\alpha_{xzzx}^{\romanI,\rmS}$, and $\alpha_{xzzx}^{\romanII,\rmS}$ are negligible.

\begin{figure}
    \hypertarget{fig3}{}
    \centering
    \includegraphics[width=1\columnwidth]{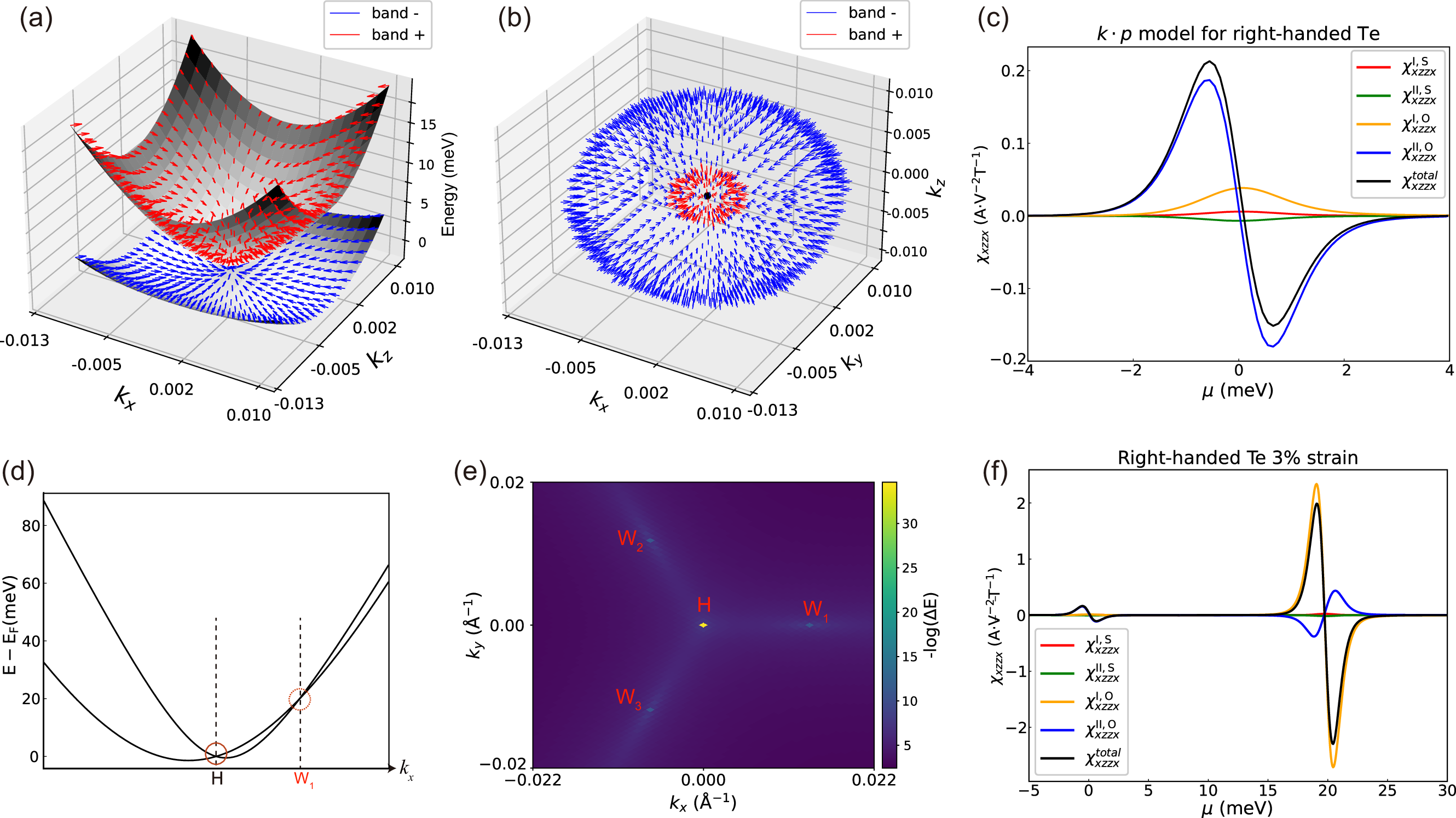}
    \caption{(a) Band structure with orbital moment (arrows) around H from the $k\cdot p$ model of right-handed Te. (b) Orbital magnetic texture in the H-centered Fermi pockets formed by the two bands ($\pm$) with $\mu$ being chosen as 2 meV above the WP. (c) $\chi_{xzzx}(\mu)$ calculated from the $k \cdot p$ model. (d) First-principles band structure along $k_x$ near H for the right-handed Te under a biaxial strain of 3\%. The solid (dashed) circle represents WP with $\mathcal{C}=+1(-1)$. (e) Local gap between two conduction bands in the $k_x$-$k_y$ plane, where three new WPs are labeled as $W_{1,2,3}$. (f) $\chi_{zxxz}$ versus $\mu$ for the 3\%-strained Te. }
    \label{fig:kpmodel}
\end{figure}

\textit{Analytic model study around WPs.}---To better understand the behavior of $\chi_{xzzx}^{\rm int}$ around WPs and its chirality dependence, we consider the following Weyl model:
\begin{equation}
\begin{split}
    H_W(\kk)={\bm h}(\kk)\cdot {\bm \sigma},
\end{split}\label{Hk.p}
\end{equation}
where ${\bm \sigma}=(\sigma_x,\sigma_y,\sigma_z)$ are Pauli matrices and ${\bm h}(\kk)=(v_1k_x,v_2k_y,v_3k_z)$ with $v_1=v_2$ constrained by the $D_{3d}$ symmetry at H.
The model describes two energy bands $\epsilon_{\pm}(\kk)=\pm h=\pm\sqrt{v_1^2k_x^2+v_2^2k_y^2+v_3^2k_z^2}$.
The orbital and spin magnetic moments are given by
\begin{equation}
    \bm{\mathcal{\bm M}}^{\pm,\rmO}=-\frac{v_1v_2v_3}{2h^2}\kk=\pm \bm{\Omega}^{\pm} h,\quad \bm{\mathcal{\bm M}}^{\pm,\rmS}=\mp g\mu_B \hat{{\bm h}}.
\end{equation}
One observes that the orbital magnetic moments $\bm{\mathcal{M}}^{\pm,\rmO}$ are the same for the two bands ($\pm$). Given that the directional vector ${\bm m}=\bm{\mathcal{M}}^{\pm,\rmO}/|\bm{\mathcal{M}}^{\pm,\rmO}|$ is along with $\kk$ [as shown in Fig.~\ref{fig:kpmodel}(a,b)], the orbital magnetism in momentum space forms a hedgehog-like texture [like Berry curvature $\bm{\Omega}(\kk)$ in Fig.~\ref{fig:band}(d)] with a winding number on the Fermi surface, $\zeta=\frac{1}{4\pi}\oint_{\rm FS} {\bm m}\cdot(\partial_\theta {\bm m}\times\partial_\phi{\bm m})ds=-{\rm sgn}(v_1v_2v_3)=\mathcal{C}$, which directly relate to the electronic chirality. However, the spin magnetic moments $\bm{\mathcal{M}}^{\pm,\rmS}$ are opposite for the two bands and only depend on the directional vector $\hat{\bm h}$, which are consistent with the first-principles results (see Fig~\ref{fig:band}(c) and SM \footnotemark[\value{footnote}]). Both spin and orbital magnetic moments are pseudovectors that reverse sign under $\bm{k}\rightarrow-\bm{k}$ in reciprocal space, i.e., $\bm{\mathcal{M}}(-\bm{k})=-\bm{\mathcal{M}}(\bm{k})$ as enforced by time-reversal symmetry. They also reverse signs in opposite chiral variants that are related by an inversion or mirror operation.

Next, we derive the geometric quantities in Eq.~\eqref{IIO},
\begin{equation}
    \alpha_{xzzx}^{\textnormal{II},{\rm O}}=\pm\frac{|v_1^3v_2v_3^3|k_x^2}{4h^7}\mathcal{C},\quad
    \alpha_{xzzx}^{\textnormal{II},\rmS}=g\mu_B\frac{v_1^3v_3^2k_x^2}{4h^6}.\label{alphaII}
\end{equation}
Generally, $\alpha^\rmO$ scales as $1/(\Delta\epsilon)^5$ and $\alpha^\rmS$ scales as $1/(\Delta\epsilon)^4$. Therefore, the orbital contribution can be more enhanced than the spin part with a decreased gap around the WP. Similarly, we derive $\alpha_{xzzx}^{\romanI}$ in Eq.~\eqref{IO} for spin and orbital (see SM \footnotemark[\value{footnote}]), and eventually obtain all contributions to the INPHE conductivity,
        $\chi_{xzzx}^{\romanI,\rmS}=-\chi_{xzzx}^{\romanII,\rmS}=\frac{g\mu_B}{24\pi^2\mu^2}\frac{|v_3|}{v_{1}}, 
\chi_{xzzx}^{\romanI,\rmO}=0$, 
and
\begin{equation}
\chi_{xzzx}^{\romanII,\rmO}=-\frac{v_3^2}{24\pi^2\mu^3}\mathcal{C}.\label{chi_k.p}
\end{equation}
Remarkably, the two spin parts have opposite signs which are canceled out, resulting in zero net contribution. Therefore, the INPHE conductivity originates primarily from part II of the orbital contribution. More importantly, $\chi_{xzzx}^{\romanII,\rmO}$ depends on the chirality $\mathcal{C}$ of the WP, implying opposite INHPE for right- and left-handed Te ($\mathcal{C}=\pm 1$). Moreover, $\chi^{\romanII,\rmO}$ would switch signs when the chemical potential is shifted from below ($\mu<0$) to above ($\mu>0$) the WP, which is consistent with the results in Fig.~\ref{fig:band}(f,g). Noticing that the quadratic terms of $\kk$, which would bend over the lower conduction band, are neglected in the above model~\eqref{Hk.p}. By including the symmetry-allowed quadratic term $H_{\rm quad}(\kk)=c_1(k_x^2+k_y^2)+c_3k_z^2$, we found that $\alpha^{\romanII,\rmO}$ remains unchanged but $\alpha^{\romanI,\rmO}$ has a nonzero value which is about one order of magnitude smaller than the former (see SM \footnotemark[\value{footnote}]). Based on $H_W+H_{\rm quad}$ with parameters fitted from first-principles band structures, we plot the conductivity in Fig.~\ref{fig:kpmodel}(c), which agrees well with Fig.~\ref{fig:band}(g).

\textit{Tilted WPs induced by strain.}---Since the Weyl fermions are fine-tunable by applying external strain \cite{PhysRevLett.114.206401, PhysRevLett.125.216402, PhysRevLett.124.136402, Niu2023high-pressure}, we further study the INPHE of Te under a biaxial strain along the hexagonal plane. As shown in Fig.~\ref{fig:kpmodel}(d,e), three titled WPs with chirality $\mathcal{C}=-1$, which are related to each other by the $C_3$ rotation, emerge around H in the 3\%-strained right-handed Te. This induces significant INHE when the Fermi level is shifted around these WPs [see Fig.~\ref{fig:kpmodel}(f)]. Remarkably, we find a considerable contribution from $\chi^{\romanI,\rmO}$ which is larger than $\chi^{\romanII,\rmO}$ and exhibit a opposite sign inversion with increasing $\mu$. Based on a tilted Weyl model $H_{\rm tilt}=H_W(\kk)+w_1k_x+w_2k_y$, we found that
\begin{equation}
    \alpha_{\rm tilt}^{\romanI,\rmO}=-\frac{w_2k_y|v_1v_2v_3^5|k_z^2}{8h^8}\mathcal{C}\label{tilt}
\end{equation}
scale as $1/(\Delta\epsilon)^5$, which is comparable with $\alpha^{\romanII,\rmO}$
(see Sec. III.B in SM \footnotemark[\value{footnote}]). By integrating Eq.~\eqref{tilt} on the Fermi surface, we estimate $\chi^{\romanI,\rmO}\propto \mathcal{C}/\mu^3$, exhibiting the similar behavior as $\chi^{\romanII,\rmO}$ in Eq.~\eqref{chi_k.p}. Therefore, $\chi^{\romanI,\rmO}$ should be strongly enhanced for tilted Weyl fermions with clear dependences on the chirality and the chemical potential, consistent with the significant peaks accompanied by a sign change as a function of $\mu$ in Fig.~\ref{fig:kpmodel}(f). Our analysis also indicates that large INPHE should be generally expected for type-II Weyl semimetals with over-tilt cone-shape band structures, such as WTe$_2$ \cite{Nat_typeIIWeyl,natphys_typeIIMoTe2}, TaIrTe$_4$ \cite{PhysRevB.93.201101,cpge_SunDong2019natmat}, and \textit{p}-doped Te \cite{PhysRevLett.114.206401, PhysRevB.95.125204, PhysRevLett.125.216402}.

\begin{figure}
    \hypertarget{3}{}
    \centering
    \includegraphics[width=1\columnwidth]{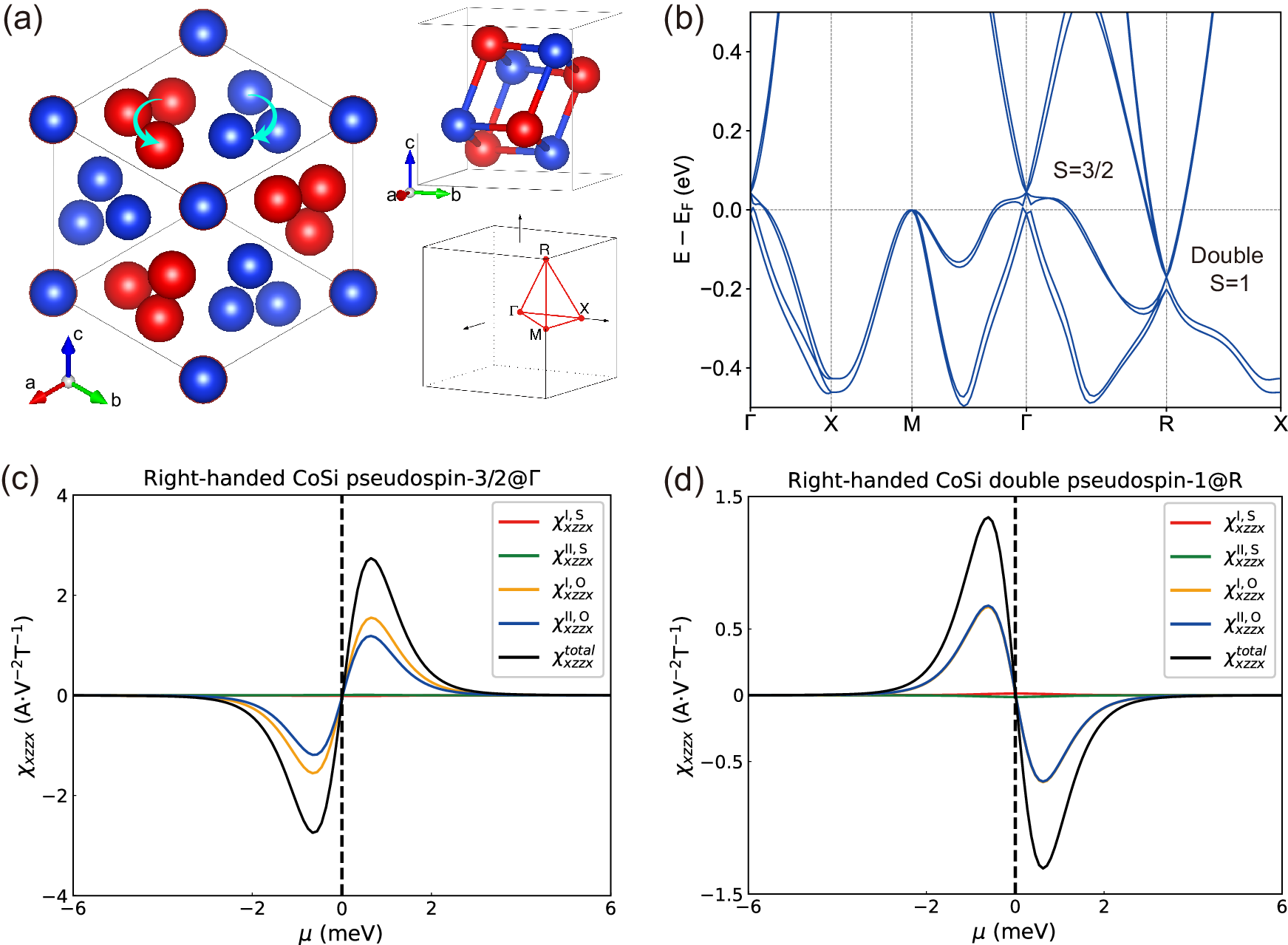}
    \caption{(a) The crystal structure of the B20 compounds AB (A=Co, Rh, Pt; B=Si, Sn, Ga, Al). The crystal is a right- or left-handed configuration when the red or blue helices represent A atoms. (b) Band structure of the right-handed CoSi where four- and six-fold degenerate nodes appear at the $\Gamma$ and R points. (c,d) $\chi_{zxxz}$ versus $\mu$ around the multifold nodes at (c) $\Gamma$ and (d) R.}
    \label{fig:main_cosi}
\end{figure}

\textit{Multifold fermions with high Chern number.}---In addition to chiral Te, we also investigate the INPHE induced by multifold fermions in topological chiral semimetals CoSi, RhSi, RhSn, PtGa, and PtAl. Figure~\ref{fig:main_cosi}(a) shows their chiral cubic structures in space group $P2_13$ (No. 198) which exhibit helices along the [111] direction. We take the right-handed CoSi as an example and defer the rest to Fig. S10-S14 in SM \footnotemark[\value{footnote}]. Figure~\ref{fig:main_cosi}(b) shows four- and six-fold degenerate nodes appear at the $\Gamma$ and R points [see Fig.~\ref{fig:main_cosi}(b)], which carry high Chern numbers of $\mathcal{C}=-4$ and 4, respectively. Given that the twofold node can be described as the spin-1/2 Weyl fermions, the four- and six-fold nodes can be regarded as spin-3/2 and double-spin-1 fermions, respectively. Based on the effective model of these multifold fermions, we analytically derived a giant INPHE that shows similar dependence on the chirality and chemical potential (See Sec. V in SM~\footnotemark[\value{footnote}]). Remarkably, we found that both $\chi^{\romanI,\rmO}$ and $\chi^{\romanII,\rmO}$ make significant contributions to the INPHE, which are proportional to their Chern numbers $\mathcal{C}$ and dominate the INPHE over spin contributions. As shown in Fig.~\ref{fig:main_cosi}(c,d), the INPHE is indeed strongly enhanced around these multifold nodes with its amplitude being about $1\sim 10\; \mathrm{A}\cdot\mathrm{V}^{-2}\cdot\mathrm{T}^{-1}$, which is $1\sim 2$ orders of magnitude larger than that of Weyl nodes.

\textit{Conclusion.}---We have shown that nodal fermions in chiral crystals can lead to prominent INPHE which depends on the electronic chirality with a compelling connection to the structural chirality of host crystals.
More importantly, we reveal that the large INPHE mainly originates from the enhanced orbital magnetism with node-centered hedgehog-like textures characterized by the topological charge. Our study reveals an exotic chirality-dependent nonlinear transport phenomenon, potentially leading to novel applications in chirality-induced nonreciprocal devices and enantioselective processes. Our work also paves a new way for studying other types of nodal fermions in more topological semimetals, which opens the possibility of exploring unprecedented physical properties of unconventional nodal fermions, such as magnetic Weyl \cite{PhysRevLett.107.186806,magneticWeyl}, Dirac \cite{PhysRevLett.108.140405,Dirac_PT_CuMnAs},
hourglass \cite{hourglass,sciadv.1602415}, triple \cite{science.aaf5037,PhysRevX.6.031003}, and other high-fold chiral fermions \cite{Diracfermion_NRM}.

\begin{acknowledgments}
We thank Prof. Cong Xiao for the valuable discussion. This work is supported by the National Key R\&D Program of China (Grant No. 2021YFA1401600) and the National Natural Science Foundation of China (Grant No. 12074006). The computational resources are supported by the high-performance computing platform of Peking University.
\end{acknowledgments}

%

\end{document}